\documentclass[aps,prl,twocolumn,showpacs]{revtex4}
\usepackage{amsmath,amssymb,graphicx,bbm}
  \graphicspath{{nerapfig/}}
  \DeclareGraphicsExtensions{.eps}

\newcommand{\Tau}{\mathrm{T}}
\newcommand{\mathd}{\mathrm{d}}
\newcommand{\tmmathbf}[1]{\ensuremath{\boldsymbol{#1}}}
\newcommand{\tmop}[1]{\ensuremath{\operatorname{#1}}}
\newcommand{\tmstrong}[1]{\textbf{#1}}
\newcommand{\tmtextit}[1]{{\itshape{#1}}}
\newcommand{\um}{-}

\begin{document}

\title{Implementing quantum gates by optimal control with doubly exponential
convergence}
\author{Pierre de Fouquieres}
\affiliation{Centre for Quantum Information and Foundations,
             Department of Applied Maths and Theoretical Physics,
             University of Cambridge, Wilberforce Road, Cambridge, 
	         CB3 0WA, United Kingdom}
\date{\today}

\begin{abstract}
  We introduce a novel algorithm for the task of coherently controlling a
  quantum mechanical system to implement any chosen unitary dynamics. It
  performs faster than existing state of the art methods by one to three
  orders of magnitude (depending on which one we compare to), particularly for
  quantum information processing purposes. This substantially enhances the
  ability to both study the control capabilities of physical systems within
  their coherence times, and constrain solutions for control tasks to lie
  within experimentally feasible regions. Natural extensions of the algorithm
  are also discussed.
\end{abstract}
\maketitle

  The coherent control of quantum mechanical systems
  {\cite{wiseman_quantum_2009,shapiro_principles_2003}} has been sucessfully
  applied to a growing number of tasks in recent years
  {\cite{dowling_quantum_2003,mabuchi_principles_2005}}. The early approaches
  such as two pathway quantum interference, pump-dump schemes, or stimulated
  Raman adiabatic passage are intrinsically understandable in terms of
  interference due to the coordinated activation of resonant transitions
  between few energy levels {\cite{brif_control_2010}}. In order to extend
  such strategies and tackle more challenging problems, the field has moved
  towards employing pulse shapers and optimization algorithms
  {\cite{singer_colloquium:_2010}}.
  
  The field has also broadened its scope, from problems of state preparation,
  or more generally maximizing the value of an observable over an ensemble
  {\cite{tannor_control_1985}}, to, in particular, implementing unitary maps
  {\cite{schulte-herbrueggen_optimal_2005}}. This latter bridges the gap
  between physical dynamics and the gate formalism of quantum information
  processing, for which high accuracy solutions are sought, ultimately aiming
  to reach an error correction threshold around $10^{- 3}$ to $10^{- 4}$ per
  gate or below {\cite{steane_overhead_2003}}. In applying the same methods to
  both state, and map or gate problems, the additional structure inherent to
  gate problems has however been neglected -- the aim of this letter is to
  describe how this structure can be exploited to better understand and
  substantially ease the solving of gate problems.
  
  Formally, a closed $N$-level system undergoes controlled unitary dynamics
  given by $U_{\text{{\tmstrong{f}}}}$ satisfying
  \[\frac{\partial}{\partial t} U_{\text{{\tmstrong{f}}}} (t) = - i H [
  \text{{\tmstrong{f}}} (t)] U_{\text{{\tmstrong{f}}}} (t), \quad
  U_{\text{{\tmstrong{f}}}} (0) = I\]
  with $I$ the identity matrix, and a time dependent Hamiltonian
  \begin{equation}
    H [ \text{{\tmstrong{f}}} (t)] = H_0 + \sum_{r = 1}^R
    \text{{\tmstrong{f}}}_r (t) H_r \label{bilin}
  \end{equation}
  where {\tmstrong{f}} is a set of $R$ controls pulses, altering the system
  potential within a semi-classical model under the bilinear approximation.
  The abstract control problem for a target unitary gate $V$ consists in
  finding a set of real valued functions {\tmstrong{f}} and an evolution time
  $T$ such that the dynamics satisfies $U_{\text{{\tmstrong{f}}}} (T) = V$.
  Since this entails transferring a full basis of states to another (with
  relative phases), the intuition applicable to state problems is no longer
  available, and in fact schemes to solve it explicitly are limited to two
  level systems, or special cases with few levels. For practical purposes, we
  would only require the actual $U_{\text{{\tmstrong{f}}}} (T)$ and target $V$
  to match up to some prescribed error level $\varepsilon$, with respect to a
  notion of distance $d$. Optimisation, whereby a sequence of pulses
  $\text{{\tmstrong{f}}}^{(n)}$ is generated iteratively with the requisite
  distance $d (U_{\text{{\tmstrong{f}}}^{(n)}} (T), V)$
  decreasing at each step, has emerged as the strategy of choice for achieving
  this. The definitions of distance $d (U, V)$ to measure the error have
  generally been based on the Hilbert-Schmidt norm, using either $\|U - V\|$
  or, quotienting out the unphysical global phase of the dynamics and
  normalising,
  \[d (U, V) = \frac{1}{2 \sqrt{N}} \min_{\varphi} \|U - e^{i \varphi} V\|\]
  which we will be using herein. In contrast to state control problems where
  intuitive understanding often plays a role in choosing the initial trail
  pulses $\text{{\tmstrong{f}}}^{(0)}$, the serious limitations of intuitive
  insight for gate problems lead to $\text{{\tmstrong{f}}}^{(0)}$ typically
  being chosen arbitrarily, eg. at random. Moreover, to get any measure of
  gate error based on experimental measurement would require exhaustive and
  arduous process tomography, so that there has been an overwhelming
  preference towards working with numerical simulation of a model for the
  system.
  
  Running a numerical optimisation algorithm requires that the control pulses
  be discretised, and in order to incorporate experimental constraints we can
  choose a basis for discretisation corresponding to the capabilities of our
  pulse shaping equipment. Thus we let
  \[\text{{\tmstrong{f}}}_r (t) = \sum_{k = 1}^K \alpha_{r k} b_k (t)\]
  and then optimise over the set of $R K$ coefficients $\alpha_{r k}$; ideally
  the basis elements $b_k$ would be precisely calibrated to the equipment, but
  for definiteness we will consider representatives of two important cases. A
  common choice in the literature, and the main one we will use, is that of
  piecewise constant functions, as can be produced by an arbitrary waveform
  generator {\cite{motzoi_optimal_2011}}. In the case of frequency domain
  pulse shaping, we are dealing with functions which, up to Gaussian tails,
  are both time and spectral bandwidth limited. To capture this property,
  under suitable scaling we can let $b_{k + 1}$ be the Hermite
  function \footnote{Also known as the eigenfunctions of the harmonic
  oscillator} of index $k$, and restrict to the first $K$ of these.
  Such a choice has on the other hand not been used in the
  quantum control setting to our knowledge, although the benefits of Hermite
  functions have certainly been exploited in other applications, eg.
  {\cite{higuchi_multiplex_1968}} -- while the smoother alternatives to 
  piecewise constant functions used, such as truncated interpolating 
  polynomials \cite{li_optimal_2011}, have much fatter tails in frequency 
  domain. In addition to the basis constraint on the pulses {\tmstrong{f}}, 
  there must clearly be some bound $B$ on the pulse fluences, equivalently 
  their magnitude in the integrated power norm.
  
    \begin{figure}[h]
      \includegraphics[width=\columnwidth]{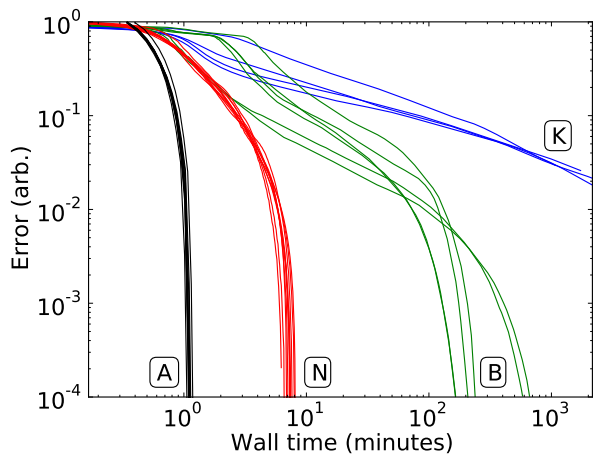}
      \caption{\label{cmp}Performance comparison between several runs of the
      Newton-Raphson (red, labeled `N'), BFGS GRAPE (green, labeled `B') and
      Krotov (blue, labeled `K') algorithms with small initial pulses
      $\text{{\tmstrong{f}}}^{(0)}$. Also
      shown (in black, labeled `A') are Newton-Raphson runs
      initialised at the norm with least ill-conditioning; the cost of finding
      this norm, on average 21 seconds, is included.}
    \end{figure}
  
  For generic intrinsic and control Hamiltonians $H_0, \ldots, H_R$ of
  (\ref{bilin}), as well as many specific cases of physical interest, full
  controllability is known to hold, meaning that any target gate can be
  achieved given sufficient evolution time $T$ and freedom in shaping the
  control pulses {\cite{jurdjevic_control_1972a}}. While this is a strong
  result, it does not identify which gates are achievable for particular
  evolution times and constraints on the controls, in any given experimental
  context. Indeed, such specific results are currently lacking, and one must
  resort to numerical investigations in order to gain better understanding
  into the capabilities of each physical system {\cite{caneva_optimal_2009}}.
  This motivates the need for an optimisation algorithm to solve the gate
  problem in runtimes on a scale rendering the process interactive or faster.
  It is the purpose of this letter to introduce a Newton-Raphson root finding
  approach for this problem, which performs substantially faster than existing
  methods (see Fig. \ref{cmp}), thereby achieving this goal on non-trivial
  examples. Such an approach also sheds light on the strong influence of the
  pulse initialisation $\text{{\tmstrong{f}}}^{(0)}$, leading to a
  prescription for how to choose it.
  
  For consistency, all numerical examples herein are for the canonical problem
  of implementing a quantum Fourier transform on a five qubit Ising coupled
  spin chain in a magnetic field gradient using two controls. But these are
  qualitatively representative of the results for different evolution times,
  target gates, and modes of control -- a different problem
  scenario illustrates such similarity in the supplement. In terms of Pauli
  matrices $\sigma$, the Hamiltonian in question is
  \begin{multline}
  H \left[ \text{{\tmstrong{f}}}  \left( t \right) \right] = \sum_{n = 1}^4
  \sigma_z^{\left( n \right)} \sigma_z^{\left( n + 1 \right)} - \sum_{n = 1}^5
  \left( n + 2 \right) \sigma_z^{\left( n \right)} \\ + \text{{\tmstrong{f}}}_1
  \left( t \right) \sum_{n = 1}^5 \sigma_x^{\left( n \right)} +
  \text{{\tmstrong{f}}}_2 \left( t \right) \sum_{n = 1}^5 \sigma_y^{\left( n
  \right)}
  \end{multline}
  while we fix an evolution time $T$ of $125$ and use $K = 1000$ basis
  functions, with piecewise constant controls unless stated otherwise.
  
  Over the last fifteen years, most techniques successfully applied to
  model-based quantum control problems have either come from mainstream
  gradient-driven optimisation theory, eg. conjugate gradient and BFGS
  {\cite{de_fouquieres_second_2011}} based GRAPE {\cite{khaneja_optimal_2005}}
  algorithms, or can be understood in this context, as with the Krotov method
  {\cite{maday_new_2003}}. These state of the art techniques have led to
  advances such as towards implementing logic gates fault-tolerantly
  {\cite{nigmatullin_implementation_2009}} or with minimal errors given the
  decoherence time {\cite{spoerl_optimal_2007}}. They owe their performance to
  the use of gradient information, but a key realisation is that the full
  Jacobian matrix $\tilde{J}$ of $U_{\text{{\tmstrong{f}}}} (T)$
  for the gate problem can be computed as efficiently as its single row
  constituting the gradient vector. Indeed the usual gradient
  computation {\cite{kuprov_derivatives_2009}}, for
  $\|U_{\text{{\tmstrong{f}}}} (T) - V\|^2$ say, effectively proceeds through
  $\tilde{J}$ by inner producting each row of $\tilde{J}$ with $V$, so that
  using the gradient alone means discarding a lot of valuable information.
  
  Looking at the singular value decomposition of the Jacobian matrix leads to
  a clean geometric picture, whereby changes to the pulses below a certain
  norm $r$ (beyond which higher order terms cease being
  negligible) induce changes in the implemented gate within a prescribed
  ellipsoid. The basic Newton-Raphson iteration
  {\cite{kelley_solving_2003}} then consists in using this Jacobian model,
  with a heuristic choice of $r$, to compute new pulses bringing the
  implemented $U (T)$ closer to the target $V$, which reduces to
  a linear algebraic task. In order to have the modelling ellipsoid strictly
  track the unitary group, one can map its elements down via the matrix
  exponential, or conversely, group elements up via the matrix logarithm, as
  we describe later. The volume of this ellipsoid determines the ability of
  all algorithms mentioned herein to shift $U_{\text{{\tmstrong{f}}}} (T)$ in
  general directions, while for the specific target $V$ a more relevant
  quantity correlated to this volume is the distance to exact
  solution controls upon ignoring higher order terms, which we shall refer to
  as the level of ill-conditioning.
  
    \begin{figure}[h]
      \includegraphics[width=\columnwidth]{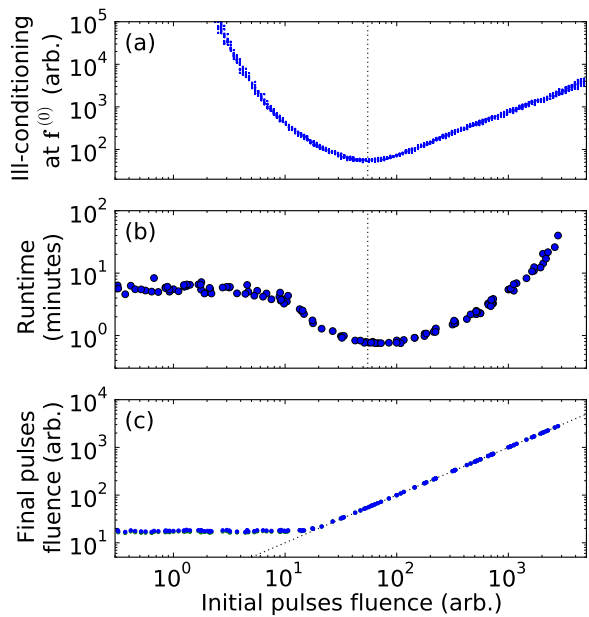}
      \caption{\label{split}For Newton-Raphson runs with initial pulses
      $\text{{\tmstrong{f}}}^{(0)}$ of different norms, (b) the
      wall time needed to reach an error $\varepsilon$ of $10^{- 4}$ and (c)
      the norm of the corresponding solution pulses, with a dashed `initial
      equals final' line. In addition, (a) the ill-conditioning of the
      Jacobian at several randomly sampled pulses of each norm.}
    \end{figure}
  
  The situation for our test problem depicted in Fig. \ref{split} is
  representative of the general structure, as described in detail in 
  the supplemental material, which can be expected of all problems. We see
  that both the ill-conditioning and the Newton-Raphson runtime strongly
  depend on the integrated power of the pulses (specifically of
  $\text{{\tmstrong{f}}}^{(0)}$ in the latter case), but are
  well concentrated beyond this along a single curve, with the minimum of
  these two curves coinciding. When solution pulses are required to have a
  fluence below $B$, we can cheaply find the minimum of the ill-conditioning
  curve restricted to the interval $\left[ 0, 4 B / 5 \right]$ say, and use an
  arbitrary $\text{{\tmstrong{f}}}^{(0)}$ with this norm to
  initialise the algorithm. It is clear from Fig. \ref{split}(c) that this
  choice will yield a solution satisfying the fluence constraint unless no
  choice can, while the correspondance between ill-conditioning and runtime
  curves makes it the most efficient choice. The benefit of this prescription
  over less deliberate ones is evident from comparing the `A' and `N' series
  of runs in Fig. \ref{cmp}.
  
  The original equation $U_{\text{{\tmstrong{f}}}} (T) = V$ should be thought
  of as over-determined when the dimension $N^2$ of the unitary group is
  greater than that of the control space $R K$, since then it is only solvable
  to arbitrarily high accuracy for an exceptional set of targets $V$. This
  makes it an unfavourable case in the context of low error control, because
  it is implausible for a target gate $V$ of interest to be special in this
  sense. On the other hand, whenever solvability is not so limited then almost
  every achievable target $V$ admits an $R K - N^2$ dimensional set of control
  pulses implementing it. Although the number of iterations for algorithms to
  reach a given error tolerance $\varepsilon$ is, as expected, reduced as this
  dimension of degeneracy increases, there is a counter-intuitive downside to
  under-determined problems.
  
\begin{figure}[h]
      \includegraphics[width=\columnwidth]{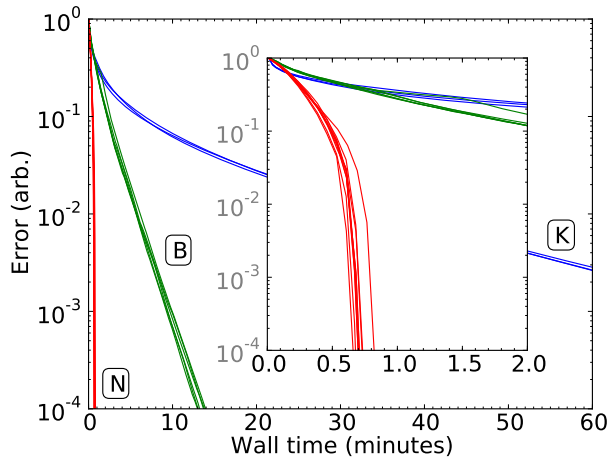}
      \caption{\label{rates}Illustration of the different convergence
      behaviours of Newton-Raphson (red, labeled `N'), BFGS GRAPE (green,
      labeled `B') and Krotov (blue, labeled `K') algorithms with initial
      norms having minimal ill-conditioning. Although several runs were
      carried out from very different initial pulses, the performance profile
      of each algorithm remains the same.}
    \end{figure}

  In general when converging to an exact solution, the error of Krotov
  iterates decays exponentially, ie. eventually as $\gamma^n$
  for some $\gamma < 1$, while with conjugate gradient or BFGS, the error
  decay is faster than exponential, and Newton algorithms have error decaying
  doubly exponentially {\cite{ben-israel_newton-raphson_1966}}, as $O (
  \beta^{2^n} )$ for some $\beta$. But in the under-determined context,
  one can only count on exponential convergence from all of
  these algorithms except for Newton-Raphson root finding which retains its
  double exponential convergence. Indeed, the directions of
  degeneracy about a solution form a null space to the Hessian there,
  rendering inapplicable the analysis {\cite{dennis_characterization_1974}} on
  which faster than exponential convergence results for BFGS are based
  {\cite{powell_global_1976,li_global_2001}}. The stark difference between
  these rates is illustrated in Fig. \ref{rates}, where all Newton-Raphson
  runs surpass $10^{- 4}$ in a single iteration once they reach $10^{- 2}$
  error.
  
  In order to make best use of Newton-Raphson root finding, we
  must re-formulate our problem over a linear space, and the most natural
  choice here is to seek for the functional
  \[\mathfrak{L}( \text{{\tmstrong{f}}}) = \log \left( V^{\dag}
  U_{\text{{\tmstrong{f}}}} (T) \right)\]
  to equal zero within the space $\mathfrak{u}$ of anti-Hermitian matrices.
  The resulting algorithm then has the elegant property of reducing the
  geodesic distance between the actual $U_{\text{{\tmstrong{f}}}} (T)$ and
  target gate $V$ on each iteration. It can usefully be made more general by
  restricting attention to a subspace of $\mathfrak{u}$, specified by an
  orthogonal projection $P$, ie. seeking a zero of $P\mathfrak{L}(
  \text{{\tmstrong{f}}})$. In particular, restricting to the space
  $\mathfrak{s}\mathfrak{u}$ of traceless anti-Hermitian matrices makes the
  root finding insensitive to the unphysical global phase.
  
  A less obvious application is to implement a gate on a system interacting
  coherently with an environment {\cite{rebentrost_optimal_2009}}, which
  contrary to Markovian interaction with a bath is reversible so need not
  fundamentally limit the achievable error. The full Hilbert space then splits
  as $S \otimes E$, and for a given gate on the system $W$ our aim would be to
  implement any gate of the form $W \otimes A$ for an ancillary evolution $A$
  -- this corresponds to letting $V = W \otimes I$ in $\mathfrak{L}$ and
  projecting it out of the space $I \otimes \mathfrak{u}$ with $P$. Although
  the Newton-Raphson algorithm can certainly be applied to Lindblad dynamics,
  choosing a short evolution time $T$ to limit dissipation and finding a
  control for the system without bath should still be a first step, since
  computing the evolution super-operator is much more expensive. In both 
  extensions, having $R K \geqslant N^2$ becomes far from sufficient to justify 
  concluding the problem is exactly solvable.
  
    \begin{figure}[h]
      \includegraphics[width=\columnwidth]{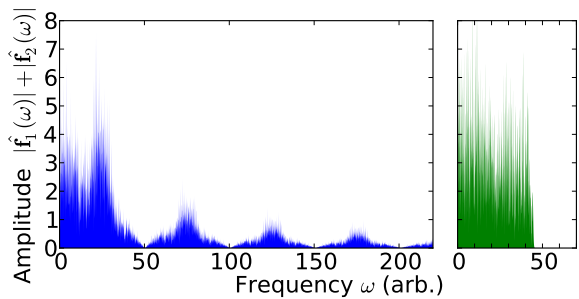}
      \caption{\label{spec}Spectrum of typical minimal norm solutions with error 
      below $10^{- 4}$, in the piecewise constant (left) and Hermite function
      (right) bases. These are symmetric about $\omega = 0$, but the Hermite
      functions could just as easily be made bandlimited about any chosen
      carrier frequency.}
    \end{figure}
  
  Up until now, all numerical examples have been in the piecewise constant
  basis, but our algorithm applies to general bases, and of particular
  interest is the Hermite basis to find spectrally narrow solution pulses.
  Note that the same bandwidth as seen in the right panel of Fig. \ref{spec}
  could be obtained by a suitable basis of low frequency Fourier components,
  but implementing such a pulse in a finite time duration would lead to
  distortion, thereby deteriorating the achieved error. Interestingly, using
  the same number of basis functions $K$, the curve from Fig. \ref{split}(c)
  and the most efficient initial fluence remain unchanged across both bases --
  the average number of iterations to reach the same error tolerance
  $\varepsilon$ is also similar (10.1 vs. 12.5) for this initial fluence. The
  piecewise constant basis is however special in how operations are cheaper
  with it than in general bases, eg. for our test problem computing the
  propagator $U_{\text{{\tmstrong{f}}}} \left( T \right)$ takes 0.9 seconds,
  as opposed to around 50 seconds in the Hermite basis. This makes it all the
  more important to choose the initial $\text{{\tmstrong{f}}}^{(0)}$
  in a general basis carefully, and to this end information from the
  more tractable piecewise constant case seems to suffice.
  
  We have seen how viewing unitary map control problems from the root finding
  perspective motivates an algorithm offering vast performance improvements
  over existing methods, and reveals particularly clean structure within the
  space of controls. This formulation can moreover naturally be made in the
  full generality of pulses represented in arbitrary bases and accounting for
  an environment to the system. For state preparation problems, considering
  the Newton-Raphson algorithm analogue {\cite{martinez_quasi-newton_1991}} in
  that case promises to lead to further fruitful developments.
  
  This work was funded by EPSRC, via CASE/CNA/07/47, and Hitachi. The author
  wishes to thank Sophie Schirmer and Peter Pemberton-Ross for valuable
  exchanges.

\section{Supplemental material}

We will now describe the algorithm as implemented for producing
numerical results in this work, aiming to include sufficient technical detail
that it may be fully replicated based on this account. This
specification focuses on the resulting performance more than ease of
implementation, and most of its prescriptions can be legitimately simplified
as needed. A source code release, to be made available as a Python
SciKit{\footnote{http://scikits.appspot.com}} under the code-name CUMIN, is 
also in preparation.

\subsection{Numerical optimisation}

A common thread in derivative-based optimisation
{\cite{nocedal_numerical_1999}}, aiming to minimise an objective function
$\mathfrak{E}: \mathbbm{R}^M \rightarrow \mathbbm{R}$, is the repeated update
of a variable $\tmmathbf{a}$ storing what can be thought of as the current
base point for the domain $\mathbbm{R}^M$. We will write $\tmmathbf{a}^{\left(
0 \right)}, \tmmathbf{a}^{\left( 1 \right)}, \ldots$ for the sequence of
values taken by $\tmmathbf{a}$, but omit the superscript when referring to its
value at some present state of the algorithm. On each iteration, say the
$n^{\tmop{th}}$, a model for $\mathfrak{E}$ valid locally about the outgoing
base point $\tmmathbf{a}^{\left( n - 1 \right)}$ for this iteration is
constructed based on the value of certain derivatives of function
$\mathfrak{E}$ at the point $\tmmathbf{a}^{\left( n - 1 \right)}$. The idea is
then to take for the next base point $\tmmathbf{a}^{\left( n \right)}$ the
minimum $\tmmathbf{v}$ of this model for $\mathfrak{E}$, possibly restricted
to some neighbourhood of $\tmmathbf{a}^{\left( n - 1 \right)}$. However, since
the model is only valid locally, we cannot guarantee $\mathfrak{E} \left(
\tmmathbf{v} \right) <\mathfrak{E}(\tmmathbf{a}^{\left( n - 1 \right)})$ ie.
that using $\tmmathbf{v}$ as the next base point will lead to a reduction of
the current objective value $\mathfrak{E} \left( \tmmathbf{a} \right)$. Two
families of approaches exist for resolving this problem: line search methods
look for a suitable point along the line starting at $\tmmathbf{a}^{\left( n -
1 \right)}$ in the direction of $\tmmathbf{v}$, while trust region methods
consider rescaling the neighbourhood of $\tmmathbf{a}^{\left( n - 1 \right)}$
to which they constrain minimisation of the model for $\mathfrak{E}$.

The main performance characteristic which can be reliably
affected in choosing between derivative-based optimisation algorithms is the
run-time required to achieve convergence to a given accuracy. When methods
have different asymptotic rates of convergence, one of them is determined in
advance to eventually be closer to the limit than all others. Prior to this
regime, there is a trade-off in the accuracy of models used on each iteration,
where more derivative information can be acquired at higher computational cost
to construct better models, which should in contrast enable greater progress
per iteration. In this respect, the Hessian matrix of all $M \left( M + 1
\right) / 2$ second derivatives is prohibitively expensive to compute except
for special $\mathfrak{E}$, so that widely effective algorithms only require
the gradient vector of $M$ first derivatives.

For example quasi-Newton optimisation methods, of which BFGS is
an instance, use the model $\mathfrak{E} \left( \tmmathbf{a}+\tmmathbf{p}
\right) \simeq \mathfrak{E} \left( \tmmathbf{a} \right) +\tmmathbf{g}^{\Tau}
\tmmathbf{p}+ \frac{1}{2} \tmmathbf{p}^{\Tau} B\tmmathbf{p}$ where
$\tmmathbf{g}$ is the gradient vector of $\mathfrak{E}$ at $\tmmathbf{a}$, and
$B$ is constructed iteratively to approximate the Hessian matrix of
$\mathfrak{E}$ at $\tmmathbf{a}$. At least for BFGS, the model minimum
$\tmmathbf{v}$, namely $- B^{- 1} \tmmathbf{g}$ since $B$ is always positive
definite, is used within a line search routine at each iteration. In this
context, steepest descent would correspond to replacing $B$ by the identity
matrix $I$, but we do not consider this method since it cannot be recommended
in general, and performs particularly slowly on gate control problems
{\cite{machnes_comparing_2011}}. For comparison, convergence of BFGS iterates
to a local minimum $\tmmathbf{a}^{\ast}$ where the Hessian of $\mathfrak{E}$
is positive definite is understood to happen at a rate, measured either by
$\|\tmmathbf{a}^{\left( n \right)} -\tmmathbf{a}^{\ast} \|$ or
$\mathfrak{E}(\tmmathbf{a}^{\left( n \right)}) -\mathfrak{E} \left(
\tmmathbf{a}^{\ast} \right)$, between $O ( \xi^{n \log n} )$
{\cite{griewank_global_1991}} and $\Omega (\xi^{n^2})$
{\cite{powell_rate_1983}} for some problem specific $\xi$. While convergence
for conjugate gradient {\cite{powell_convergence_1976}} and limited memory
BFGS {\cite{liu_limited_1989}}, without being restarted every $M$ iterations
only goes as $\xi^n$ asymptotically, which is also the rate for steepest
descent albeit with $\xi$ deteriorating significantly
{\cite{akaike_successive_1959}} on poorly conditioned problems.

\subsection{Newton-Raphson method}

This description can be adapted to cover Newton-Raphson root finding applied
to the vector valued function $\mathfrak{L}: \mathbbm{R}^M \rightarrow
\mathbbm{R}^m$ by introducing an objective function $\mathfrak{E} \left(
\tmmathbf{x} \right) =\|\mathfrak{L} \left( \tmmathbf{x} \right) \|^2$ and
specifying that the model, hence the derivatives used to construct it, should
be of $\mathfrak{L}$ rather than $\mathfrak{E}$. The model in this case reads
\[\mathfrak{L} \left( \tmmathbf{a}+\tmmathbf{p} \right) \simeq \mathfrak{L}
\left( \tmmathbf{a} \right) + J\tmmathbf{p}\]
where $J$ is the Jacobian matrix of $\mathfrak{L}$ at the point
$\tmmathbf{a}$, with entry $i, j$ equal to the $j^{\tmop{th}}$ partial
derivative $\frac{\partial}{\partial x_j}$ of the $i^{\tmop{th}}$ component of
$\mathfrak{L} \left( \tmmathbf{x} \right)$ -- in the notation of differential
geometry, we can succinctly write $J$ as $\mathd
\mathfrak{L}|_{\tmmathbf{a}}$.

If the model were exact, assuming the rank of $J$ equals $m$, a global minimum
of $\mathfrak{E}$ could be found where $\mathfrak{L}$ is $\tmmathbf{0}$,
namely at $\tmmathbf{a}+\tmmathbf{p}$ with $\tmmathbf{p}$ being any solution
to $J\tmmathbf{p}= -\mathfrak{L} \left( \tmmathbf{a} \right)$. Indeed the
classical Newton-Raphson algorithm, applicable for dimensions $M = m$, uses
the update rule $\tmmathbf{a}^{\left( n + 1 \right)} =\tmmathbf{a}^{\left( n
\right)} +\tmmathbf{p}$ for which $\mathfrak{E}(\tmmathbf{a}^{\left( n
\right)})$ goes to zero doubly exponentially in $n$, specifically as $O (
\beta^{2^n} )$ for some $\beta$ (known as quadratic convergence), for
suitable initial conditions $\tmmathbf{a}^{\left( 0 \right)}$. Generalising
this to the under-determined regime $M > m$, the value of $\tmmathbf{p}$ is no
longer explicitly determined as $\um J^{- 1} \mathfrak{L} \left( \tmmathbf{a}
\right)$, and a suitable choice for $\tmmathbf{p}$ which preserves the
quadratic convergence property is the minimum norm solution of $J\tmmathbf{p}=
-\mathfrak{L} \left( \tmmathbf{a} \right)$
{\cite{ben-israel_newton-raphson_1966_sup}}, computable as $\um J^{\Tau} (J
J^{\Tau})^{- 1} \mathfrak{L} \left( \tmmathbf{a} \right)$. In either case, a
suitable initial $\tmmathbf{a}^{\left( 0 \right)}$ would be any point
sufficiently close to some solution $\tmmathbf{a}^{\ast}$ of $\mathfrak{L}
\left( \tmmathbf{a}^{\ast} \right) =\tmmathbf{0}$ where the Jacobian $\mathd
\mathfrak{L}|_{\tmmathbf{a}^{\ast}}$ has full rank $m$, in such a way that
$\mathd \mathfrak{L}|_{\tmmathbf{a}^{\left( n \right)}}$ is always full rank
so that all iterates $\tmmathbf{a}^{\left( n \right)}$ are well-defined.

Finding such an initial point is itself difficult, and for general points
$\tmmathbf{a}$ the neighbourhood in which the model at $\tmmathbf{a}$ is
accurate contains neither a root $\tmmathbf{a}^{\ast}$ of $\mathfrak{L}$, nor
a root $\tmmathbf{a}+\tmmathbf{p}$ of the model. The rationale behind the
update rule $\tmmathbf{a}^{\left( n + 1 \right)} =\tmmathbf{a}^{\left( n
\right)} +\tmmathbf{p}$, that $\tmmathbf{a}^{\left( n + 1 \right)}$ should be
close to a root of $\mathfrak{L}$ since $\mathfrak{L}(\tmmathbf{a}^{\left( n
\right)} +\tmmathbf{p}) \approx \tmmathbf{0}$, therefore no longer applies for
general $\tmmathbf{a}^{\left( n \right)}$ and we are compelled to invoke a
line search or trust region method to find a usable $\tmmathbf{a}^{\left( n +
1 \right)}$. But before we delve further into this, let us deal more
specifically with the function $\mathfrak{L}$ which we use for the unitary map
control problem. Note in passing that for a general objective
$\mathfrak{E}$ which is strictly convex at some minimiser
$\tmmathbf{a}^{\ast}$, applying Newton-Raphson to its gradient vector
$\tmmathbf{g} \left( \tmmathbf{x} \right) = \mathd
\mathfrak{E}|_{\tmmathbf{x}}$ will seek a critical point of $\mathfrak{E}$ so
must converge to $\tmmathbf{a}^{\ast}$ when starting sufficiently close to it.
The direction of line searches would then be $- J^{- 1} \tmmathbf{g}$ where
$J$ is the Hessian matrix of $\mathfrak{E}$, justifying the approximating $-
B^{- 1} \tmmathbf{g}$ used in BFGS.

\subsection{Unitary map problem}

Consider the dynamical Lie algebra $\mathfrak{l}$ of our control system, ie.
the linear space spanned by all iterated commutator expressions starting with
the matrices $i H_0, i H_1, \ldots, i H_R$. By the Frobenius theorem, the
propagators $U_{\text{{\tmstrong{f}}}} \left( t \right)$ must remain within
the associated group $e^{\mathfrak{l}}$, consisting of matrix exponentials of
matrices in $\mathfrak{l}$, for all time and over all possible control vectors
$\left( f_1, \ldots, f_R \right)$. Conversely, the controllability theory of
bi-linear systems shows that for compact $e^{\mathfrak{l}}$ there is a
critical time $T_c$ depending only on the system such that every point of
$e^{\mathfrak{l}}$ is accessible in some time $T < T_c$ using some control
vector {\tmstrong{f}}. This result also holds if the class of admissible
controls is restricted to both smooth or piecewise constant functions. At
least when in addition $\mathfrak{l}$ is semi-simple
{\cite{helgason_differential_1981}}, the set of unitary matrices
$U_{\text{{\tmstrong{f}}}} \left( T \right)$ which are accessible at a fixed
time $T$ using some control vector {\tmstrong{f}} has non-empty interior
within $e^{\mathfrak{l}}$ {\cite{jurdjevic_control_1972b}}, for each $T > 0$,
and in fact equals $e^{\mathfrak{l}}$ for $T$ sufficiently large. The
conditions of $e^{\mathfrak{l}}$ being compact and $\mathfrak{l}$ semi-simple
will be assumed in what follows -- they hold in particular for the Lie algebra
of traceless $N \times N$ anti-Hermitian matrices $\mathfrak{s}\mathfrak{u}
\left( N \right)$, which is of special interest as it corresponds to full
controllability up to global phase.

Once we have introduced a desired parametrisation of the controls, the
discretised propagator $U_{\tmmathbf{a}} \left( T \right)$ is a function
mapping $\mathbbm{R}^{R K} \rightarrow \text{U} \left( N \right)$, taking a
vector $\tmmathbf{a}$ composed of all parameters $\alpha_{r k}$ to the unitary
matrix describing the evolution under the corresponding controls. Explicitly,
$U_{\tmmathbf{a}} \left( T \right)$ equals the functional
$U_{\text{{\tmstrong{f}}}} \left( T \right)$ composed with the synthesis
operator taking $\tmmathbf{a}$ to the vector {\tmstrong{f}} of functions such
that $\text{{\tmstrong{f}}}_r (t) = \sum_{k = 1}^K \alpha_{r k} b_k (t)$.
Solving $U_{\tmmathbf{a}} \left( T \right) = V$ can naturally be phrased as
finding the root of either $U_{\tmmathbf{a}} \left( T \right) - V$ or
$V^{\dag} U_{\tmmathbf{a}} \left( T \right) - I$, but both of these functions
range over a non-linear space, making them unsuitable for use in the
Newton-Raphson algorithm.

Now let $\mathfrak{u} \left( N \right)$ denote the real linear space of all $N
\times N$ anti-Hermitian matrices with inner product $\langle A, B \rangle =
\tmop{Tr} \left( A^{\dag} B \right)$. Amongst maps taking the unitary group to
a linear space, the inverse $\log : \text{U} \left( N \right) \rightarrow
\mathfrak{u} \left( N \right)$ of the well-studied exponential map is a
complex analytic function whose value at $I$ is $0$ and derivative there is
the identity on $\mathfrak{u} \left( N \right)$ -- for $N > 2$, it is
crucially the only such function taking the global phase neglecting subgroup
$\tmop{SU} \left( N \right)$ to a linear space
{\cite{kang_volume-preserving_1995}}. The most natural way to linearise our
problem is therefore to look for a root of
\[\mathfrak{L} \left( \tmmathbf{a} \right) = P \log \left( V^{\dag}
U_{\tmmathbf{a}} \left( T \right) \right)\]
where the branches of the logarithm are chosen to give a result in
$\mathfrak{l}+ i\mathbbm{R}I$ of minimal norm and $P$ then expresses this in
an orthonormal basis of $\mathfrak{l}$, dropping any $i I$ component. Note
that when $\mathfrak{l}$ is $\mathfrak{s}\mathfrak{u} \left( N \right)$, the
choice of branches for each eigenvalue is the standard one with imaginary part
in $\left( \um \pi, \pi \right]$, while the implementation of $P$ can just
keep each strictly upper triangular entry of the input matrix and expresses
the imaginary part of its diagonal in any orthonormal basis containing the
vector of all $1 / \sqrt{N}$. This choice for $\mathfrak{L}$ also has the
advantage, when $e^{\mathfrak{l}}$ is compact, of making the corresponding
error $\sqrt{\mathfrak{E} \left( \tmmathbf{a} \right)} =\|\mathfrak{L} \left(
\tmmathbf{a} \right) \|$ equal the geodesic distance between $U_{\tmmathbf{a}}
\left( T \right)$ and $V$ over the group $e^{\mathfrak{l}} / \text{U} \left( 1
\right)$, ie. $e^{\mathfrak{l}}$ quotiented out by global phase (see
{\cite{arvanitogeorgos_introduction_2003}} Sect. 3.2).

In its strong form {\cite{federer_geometric_1996}}, Sard's theorem gives that
the set of target gates $V$ in $e^{\mathfrak{l}} / \text{U} \left( 1 \right)$
for which a solution $\tmmathbf{a}^{\ast}$ to $U_{\tmmathbf{a}^{\ast}} \left(
T \right) \equiv V$ exists with rank deficient Jacobian $\mathd
\mathfrak{L}|_{\tmmathbf{a}^{\ast}}$ is of Hausdorff dimension at most $m -
1$. Here $m$ is still the dimension of the co-domain of $\mathfrak{L}$, namely
$\mathfrak{l}$, which makes $m = N^2 - 1$ in the most prominent case when
$\mathfrak{l}$ is $\mathfrak{s}\mathfrak{u} \left( N \right)$. Suppose the
$K$, with $M = R K > m$, linearly independent basis functions $b_k$ are not so
degenerate that the image of $\mathfrak{L}$ is a measure zero set. With
respect to choosing a target $V$ from the accessible set, of all
$U_{\tmmathbf{a}} \left( T \right)$ for fixed $T$, the full rank condition
implying quadratic convergence of the Newton-Raphson algorithm therefore
occurs with probability one.

\subsection{Computing the Jacobian}

The chain rule gives a decomposition for each column
\begin{equation}
  \frac{\partial}{\partial \alpha_{r k}} \mathfrak{L} \left( \tmmathbf{a}
  \right) = P \mathd \log |_{V^{\dag} U_{\tmmathbf{a}} \left( T \right)}
  \left( V^{\dag}  \frac{\partial}{\partial \alpha_{r k}} U_{\tmmathbf{a}}
  \left( T \right) \right) \label{col}
\end{equation}
of the Jacobian of $\mathfrak{L}$, with the derivative of the discretised
propagator known to be
\begin{equation}
  \frac{\partial}{\partial \alpha_{r k}} U_{\tmmathbf{a}} \left( T \right) =
  \um \int_0^T U_{\tmmathbf{a}} \left( T \right) U_{\tmmathbf{a}} \left( t
  \right)^{\dag} i H_r b_k \left( t \right) U_{\tmmathbf{a}} \left( t \right)
  \mathd t \label{deriv}
\end{equation}
Writing $W = V^{\dag} U_{\tmmathbf{a}} \left( T \right)$, given that the
composition $\exp \circ \log$ is the identity on $\text{U} \left( N \right)$,
the $\mathd \log |_W$ part of (\ref{col}) is simply $\left( \mathd \exp
|_{\log \left( W \right)} \right)^{- 1}$, which admits a closed form
expression. Indeed, for a general anti-Hermitian matrix $A$, if $\lambda_r$
are the eigenvalues of $A$ and $\Lambda$ a corresponding matrix of
eigenvectors (one per column), $\mathd \exp |_A \left( D \right)$ can be
computed as $e^A \Lambda \left( \Gamma \cdot \left( \Lambda^{\dag} D \Lambda
\right) \right) \Lambda^{\dag}$. Here the dot denotes elementwise
multiplication and $\Gamma$ is the matrix with entries $\Gamma_{r s} = \gamma
\left( \lambda_s - \lambda_r \right)$ where $\gamma$ is the function $z
\mapsto \frac{e^z - 1}{z}$ continuously extended, so that $\gamma \left( 0
\right) = 1$. So $\mathd \log |_W \left( B \right)$ is $\Lambda \left( \left(
\Lambda^{\dag} W^{\dag} B \Lambda \right) / \Gamma \right) \Lambda^{\dag}$
where division is carried out elementwise and the $\lambda_r$ and $\Lambda$
are those from the eigen-decomposition of $\log \left( W \right)$.

For computing the propagator $U_{\tmmathbf{a}} \left( T \right)$, we use a
fixed time stepping scheme which for $s = 1, \ldots, S$ successively evaluates
the two point propagator $U_{\tmmathbf{a}} \left( t_s, t_{s - 1} \right)$,
defined as $U_{\tmmathbf{a}} \left( t_s \right) U_{\tmmathbf{a}} \left( t_{s -
1} \right)^{\dag}$, numerically where $t_0 = 0$, $t_S = T$, and $t_s - t_{s -
1}$ is the same for each $s$. In the piecewise constant control
parametrisation case, we let $S = K$ so that all the $U_{\tmmathbf{a}} \left(
t_s, t_{s - 1} \right)$ are matrix exponentials -- computing the
eigen-decomposition of the constant Hamiltonian over each time interval
$\left( t_s, t_{s - 1} \right)$ then renders the computation of both the
propagators and their derivatives relatively inexpensive. In other cases, such
as our Hermite function parametrisation, any general ODE solver could be used,
but we found the Magnus-4 method {\cite{iserles_time_2001}} which is
specialised for linear ODEs to be accurate for a smaller number of steps $S$
than in particular the standard Runge-Kutta method, also of fourth order. To
evaluate the integral from (\ref{deriv}), given that we know
$U_{\tmmathbf{a}}$ at the endpoints of each interval $\left( t_{s - 1}, t_s
\right)$ a Lobatto quadrature rule is most appropriate, particularly the
fourth order rule in order to match the accuracy of the propagator
$U_{\tmmathbf{a}} \left( t \right)$. This rule would approximate an integral
$\int_{t_{s - 1}}^{t_s} g \left( \tau \right) \mathd \tau$ by the weighted sum
\[\frac{1}{t_s - t_{s - 1}} \left[ \frac{1}{6}\, g ( t_{s - 1} ) +
\frac{2}{3}\, g ( \left( t_{s - 1} + t_s \right) / 2 ) + \frac{1}{6}\,
g ( t_s ) \right]\]
so that for the whole $\int_0^T g \left( \tau \right) \mathd \tau$ each 
value $g \left( t_1 \right), \ldots, g \left( t_{S - 1} \right)$ enters
twice. We also used polynomial interpolation, based on the value and
first derivative at $t_{s - 1}$ and $t_s$, to evaluate $U_{\tmmathbf{a}}$ at
the midpoint of each interval $\left( t_{s - 1}, t_s \right)$, together with
precomputed values for each $b_k$ at the full set of quadrature nodes $t_s$
and $\left( t_{s - 1} + t_s \right) / 2$.

\subsection{Trust-region approach}

As described earlier, as soon as $\tmmathbf{a}$ is sufficiently close to a
solution we should update $\tmmathbf{a}$ to $\tmmathbf{a}+\tmmathbf{p}$, where
$\tmmathbf{p}$ is a solution to $J\tmmathbf{p}= -\mathfrak{L} \left(
\tmmathbf{a} \right)$, on all subsequent iterations for which the Jacobian $J
= \mathd \mathfrak{L}|_{\tmmathbf{a}}$ is full rank. However, we need an
update rule which is effective in general, and for this purpose we have found
in practise that far from a solution the trust region method readily delivers
larger decreases in square error $\mathfrak{E}$ than doing a line search
could, so for this reason we focus on the former.

The trust region approach to the Newton-Raphson method consists in finding
\begin{equation}
  \tmmathbf{p}= \underset{\tmmathbf{x}: \|\tmmathbf{x}\| \leqslant
  r}{\tmop{argmin}} \|J\tmmathbf{x}+\mathfrak{L} \left( \tmmathbf{a} \right)
  \|^2 \label{nrsp}
\end{equation}
for some trust region radius $r$, with the situation when the unconstrained
minimum can be achieved requiring that we find the solution to $J\tmmathbf{x}=
\um \mathfrak{L} \left( \tmmathbf{a} \right)$ having smallest norm. In the
over-determined case $m > R K$, independently of whether the original root
finding problem is solvable, the equation $J\tmmathbf{x}= \um \mathfrak{L}
\left( \tmmathbf{a} \right)$ generically admits no solution so that minimising
the norm of the residual $J\tmmathbf{x}+\mathfrak{L} \left( \tmmathbf{a}
\right)$ is the best we can aim for. In this case one would solve the
optimisation problem (\ref{nrsp}) directly as an instance
\[\underset{\tmmathbf{x}: \|\tmmathbf{x}\| \leqslant r}{\tmop{argmin}}\;
\tmmathbf{x}^{\Tau} J^{\Tau} J\tmmathbf{x}+ 2\mathfrak{L} \left( \tmmathbf{a}
\right)^{\Tau} J\tmmathbf{x}\]
of the so called trust region sub-problem, which consists in minimising a
quadratic function $\tmmathbf{x}^{\Tau} A\tmmathbf{x}+ 2\tmmathbf{g}^{\Tau}
\tmmathbf{x}$ over a ball of radius $r$. Without loss of generality $A$ can be
taken symmetric and the ball centred at the origin, then for $A$ semi-definite
as in our situation a solution to the sub-problem can always be found of the
form $\tmmathbf{x}_{\lambda} = - \left( A - \lambda I \right)^{- 1}
\tmmathbf{g}$ for some $\lambda \leqslant 0$. To find the appropriate
$\lambda$, we use the higher order analogue of the method from
{\cite{more_computing_1983}}, although for ease of implementation a general
convex optimisation package such as CVXOPT could be invoked to solve the
sub-problem itself.

Otherwise when $m < R K$, we can restrict attention to $\tmmathbf{x}$
orthogonal to the null space of $J$ by expressing it as $\tmmathbf{x}=
J^{\Tau} \tmmathbf{y}$, reducing the problem to
\begin{multline}
\underset{\tmmathbf{y}: \|J^{\Tau} \tmmathbf{y}\| \leqslant r}{\tmop{argmin}}
\|J J^{\Tau} \tmmathbf{y}+\mathfrak{L} \left( \tmmathbf{a} \right) \|^2 = \\
\underset{\tmmathbf{y}: \tmmathbf{y}^{\Tau} M\tmmathbf{y} \leqslant
r^2}{\tmop{argmin}} \tmmathbf{y}^{\Tau} M^2 \tmmathbf{y}+ 2\mathfrak{L} \left(
\tmmathbf{a} \right)^{\Tau} M\tmmathbf{y}
\end{multline}
where the matrix $M = J J^{\Tau}$ is defined on the co-domain of
$\mathfrak{L}$. This last problem can be solved through the trust region
sub-problem instance
\[\underset{\tmmathbf{z}: \|\tmmathbf{z}\| \leqslant r}{\tmop{argmin}}\;
\tmmathbf{z}^{\Tau} M\tmmathbf{z}+ 2\mathfrak{L} \left( \tmmathbf{a}
\right)^{\Tau} \sqrt{M} \tmmathbf{z}\]
then using any solution of $\sqrt{M} \tmmathbf{y}=\tmmathbf{z}$, all of them
being equivalent in that they yield the same final $\tmmathbf{x}$. Such a
reformulation is advantageous since it makes the corresponding matrix $A$
lower dimensional, and importantly for performance, instances of $\sqrt{M}$ in
the algorithm we use appear in such a way that it never needs to be computed.

In adapting the choice of trust region radius $r$, we strive on each iteration
to use the value $r_0$ of $r$ for which $\mathfrak{E} \left(
\tmmathbf{a}+\tmmathbf{p}_r \right)$ attains its minimum, where
$\tmmathbf{p}_r$ is the solution from (\ref{nrsp}) with radius $r$. Note that
by definition any choice of radius $r$ above the norm $r^{\ast}$ of the least
square solution to $J\tmmathbf{x}= \um \mathfrak{L} \left( \tmmathbf{a}
\right)$ is equivalent, moreover along $\tmmathbf{a}+\tmmathbf{p}_r$ the model
for $\mathfrak{E}$, namely $\|\mathfrak{L} \left( \tmmathbf{a} \right) +
J\tmmathbf{p}_r \|^2$, is strictly decreasing as $r$ ranges over $\left[ 0,
r^{\ast} \right]$. As long as the model for $\mathfrak{L} \left(
\tmmathbf{a}+\tmmathbf{p}_r \right)$ is accurate, the true value of
$\mathfrak{E}$ along $\tmmathbf{a}+\tmmathbf{p}_r$ must track its model value
and therefore be decreasing -- but once $r$ is large enough that the model for
$\mathfrak{L}$ starts to break down in the vicinity of
$\tmmathbf{a}+\tmmathbf{p}_r$, the increments $\tmmathbf{p}_{r + \delta}
-\tmmathbf{p}_r$ quickly become meaningless, making them overwhelmingly likely
to lead the true $\mathfrak{E} \left( \tmmathbf{a}+\tmmathbf{p}_r \right)$ to
increase. This causes the relative error of the model for $\mathfrak{E} \left(
\tmmathbf{a}+\tmmathbf{p}_r \right) -\mathfrak{E} \left( \tmmathbf{a} \right)$
to undergo a swift transition from small, in fact vanishing as $r \rightarrow
0$, to large magnitudes as $r$ grows past the minimiser $r_0$. In our
implementation, we adjusted $r$ to make this relative error satisfy
\[0.2 \leqslant - \frac{\|\mathfrak{L} \left( \tmmathbf{a} \right) +
J\tmmathbf{p}_r \|^2 -\mathfrak{E} \left( \tmmathbf{a} \right)}{|\mathfrak{E}
\left( \tmmathbf{a}+\tmmathbf{p}_r \right) -\mathfrak{E} \left( \tmmathbf{a}
\right) |} \leqslant 0.3\]
which typically places $r$ close to $r_0$, although the choice is a valid one
irrespectively (see {\cite{nocedal_numerical_1999}} Sect. 4.0).

\subsection{Norm dependent structure\label{nodesec}}

At the null control vector ie. $\tmmathbf{a}=\tmmathbf{0}$, the propagator
$U_{\tmmathbf{a}} \left( t \right)$ reduces to the matrix exponential $e^{- i
H_0 t}$, so that working in an eigenbasis of $H_0$ it is easy to see that
within any $U_{\tmmathbf{a}} \left( t \right)^{\dag} H_r U_{\tmmathbf{a}}
\left( t \right)$ expression from (\ref{deriv}), each diagonal entry will be a
constant function. Therefore over $\mathfrak{s}\mathfrak{u} \left( N \right)$,
only $R$ out of $N - 1$ possible linear combinations of diagonal entries can
be generated by any integral $\int_0^T U_{\tmmathbf{a}} \left( t
\right)^{\dag} H_r b \left( t \right) U_{\tmmathbf{a}} \left( t \right) \mathd
t$ where $b$ is a scalar valued function. Hence the Jacobian at null controls
of $U_{\tmmathbf{a}} \left( t \right)$, thus also of $\mathfrak{L}$, will be
rank deficient for non-trivial systems, since it would be unrealistic for a
system to have $R \geqslant N - 1$ controls unless it were of very low
dimension $N$. Then the model for $\mathfrak{L}$ at
$\tmmathbf{a}=\tmmathbf{0}$ almost surely admits no exact solution, so that
the ill-conditioning defined as $\|J^{\Tau} (J J^{\Tau})^{- 1}
\mathfrak{L} \left( \tmmathbf{a} \right) \|$ is effectively infinite there,
and by continuity tends to infinity as $\|\tmmathbf{a}\| \rightarrow 0$.

At the other extreme, {\tmstrong{f}} being large introduces high frequency
oscillation in $U_{\text{{\tmstrong{f}}}} \left( t \right)$ and
$U_{\text{{\tmstrong{f}}}} \left( t \right)^{\dag} H_r
U_{\text{{\tmstrong{f}}}} \left( t \right)$, which cancels out when
integrating $U_{\text{{\tmstrong{f}}}} \left( t \right)^{\dag} H_r b_k \left(
t \right) U_{\text{{\tmstrong{f}}}} \left( t \right)$ for any fixed basis
elements $b_k$. In other words, eigenfunctions of the infinite dimensional
Jacobian $\mathd U_{\text{{\tmstrong{f}}}} \left( T \right)$ have a lot of
their spectrum in high Fourier components, and this is lost when restricting
to the lower frequency subspace spanned by the $b_k$ to obtain the finite
dimensional $J$. As a consequence the singular values of the discretised
Jacobian $J$ shrink as $\|\tmmathbf{a}\|$ grows, with the corresponding
ill-conditioning almost surely going to infinity. This explains why the
ill-conditioning curve from Fig. 2(a) grows towards both small and
large norms, thereby attaining its minimum at some finite norm value.

For any given target gate $V$, there is a minimal norm $\mu$ below which no
solution in the chosen basis can be found to the control problem, up to the
tolerated error. Due to the high dimensionality $R K$ of the discretised
control space, the volume of parameter vectors $\tmmathbf{a}$ below some norm
$x$ increases extremely quickly with $x$, eg. for our test problem increasing
$x$ by $10\%$ will make the volume grow by a factor of over $10^{82}$. While
it is tautological that any successful algorithm run must terminate with
$\|\tmmathbf{a}\|$ above $\mu$, there should be no shortage of solutions with
norms slightly above $\mu$, so we can expect the final norm to be close to
$\mu$ when starting with $\|\tmmathbf{a}^{\left( 0 \right)} \|< \mu$. When the
norm of any iterate $\tmmathbf{a}^{\left( n \right)}$ is above $\mu$, it is
reasonable to assume the update $\tmmathbf{a}^{\left( n + 1 \right)}
-\tmmathbf{a}^{\left( n \right)}$ has no preferred direction, which by the
high dimensionality would imply it is near orthogonal to $\tmmathbf{a}^{\left(
n \right)}$ with high probability. Since the trust region radius $r$, hence
the update, should be noticeably smaller than the current iterate
$\tmmathbf{a}^{\left( n \right)}$ by a factor $1 / \rho \gg 1$ say, we would
conclude that $\|\tmmathbf{a}^{\left( n + 1 \right)} \|$ is only a factor of
$1 + \rho^2 / 2$ greater than $\|\tmmathbf{a}^{\left( n \right)} \|$.
Therefore when starting with $\|\tmmathbf{a}^{\left( 0 \right)} \|> \mu$, none
of the algorithm iterations change the norm of $\tmmathbf{a}$ substantially,
making the final norm close to the initial norm. These considerations account
for the characteristic shape of the curve in Fig. 2(c), which
matches the identity function down to some floor level, presumably equal to
$\mu$, below which it hovers just above the floor level.

Given that the ill-conditioning measures how difficult it is to decrease the
error on a given iteration, when $\|\tmmathbf{a}^{\left( 0 \right)} \|> \mu$
and all iterates $\tmmathbf{a}$ have the same norm, runs of the algorithm are
faster if and only if the ill-conditioning is lower for this norm. Otherwise
the norm of iterates $\tmmathbf{a}$ increases up to some value above $\mu$,
but since the ill-conditioning curve is increasing below $\mu$, lower initial
norms lead to runs being slower. The runtime does not however blow up as
$\|\tmmathbf{a}^{\left( 0 \right)} \| \rightarrow 0$ because even starting
from $\tmmathbf{0}$, although the first iteration may only reduce the error by
a negligible amount, the resulting $\|\tmmathbf{a}^{\left( 1 \right)} \|$
equal to the trust region radius will be substantial. Finally, this explains
the correspondence in Fig. 2 between the runtime curve and
ill-conditioning curve.

\subsection{Test problem}

The system used for numerical illustrations in this letter was a chain of five
qubits with nearest neighbour Ising coupling, with a linear gradient
inhomogeneity in the magnetic field to enable some degree of frequency
selective addressing. Explicitly, the intrinsic Hamiltonian is
\[H_0 = \sum_{n = 1}^4 \sigma_z^{\left( n \right)} \sigma_z^{\left( n + 1
\right)} - \sum_{n = 1}^5 \omega_n \sigma_z^{\left( n \right)}\]
with frequencies $\omega_n = n + 2$, and where $\sigma_z^{\left( n \right)}$
is the Pauli $z$ matrix acting on the $n^{\tmop{th}}$ spin, eg.
$\sigma_z^{\left( 2 \right)} = I \otimes \sigma_z \otimes I \otimes I \otimes
I$. The control Hamiltonians are
\[H_1 = \sum_{n = 1}^5 \sigma_x^{\left( n \right)}, \quad H_2 = \sum_{n = 1}^5
\sigma_y^{\left( n \right)}\]
corresponding to the $x$-coordinate and $y$-coordinate components
$\text{{\tmstrong{f}}}_1$ and $\text{{\tmstrong{f}}}_2$ of an electric pulse
applied simultaneously to all qubits. For this problem, the total evolution
time is fixed at $T = 125$, and the number of basis functions (per control) is
chosen to be $K = 1000$, with all data in the first three figures coming from using
piecewise constant controls. By piecewise constant, we formally mean that $b_1
(t)$ is vanishing for $t$ outside the interval $(0, T / K)$ and constant equal
to one inside, with each $b_k$ equal to the previous $b_{k - 1}$ translated
forward in time by $T / K$. The Hermite functions refer to the eigenfunctions
of the quantum harmonic oscillator, shifted to be centred at $T / 2$, and
jointly scaled about $T / 2$ so that the maximum any of the first $K$ attain
outside $\left( 0, T \right)$ is $10^{- 8}$. Moreover, the definition of
integrated power norm for the control vector {\tmstrong{f}} we use satisfies
\[\| \text{{\tmstrong{f}}} \|^2= \sum_{r = 1}^R \int_0^T |
\text{{\tmstrong{f}}}_r \left( t \right) |^2 \mathd t \]
which equals the standard Euclidian norm of the parameter vector
$\tmmathbf{a}$ when the basis functions $b_1, \ldots, b_K$ are orthonormal.

\begin{figure}[ht]
  \includegraphics[width=\columnwidth]{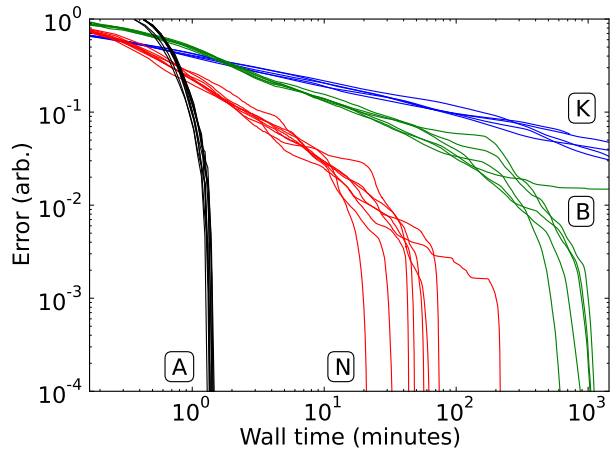}
  \caption{Corresponding
  to Fig. 1 from the main text, here for our second test problem. Performance
  comparison between several runs of the Newton-Raphson (red, labeled `N'), BFGS
  GRAPE (green, labeled `B') and Krotov (blue, labeled `K') algorithms with
  moderately sized initial pulses $\| \text{{\tmstrong{f}}}^{ \left( 0 \right)}
  \|= 20$. Also shown (in black, labeled `A') are Newton-Raphson runs preceded
  by a routine to find the norm with least ill-conditioning.}
\end{figure}

\begin{figure}[ht]
  \includegraphics[width=\columnwidth]{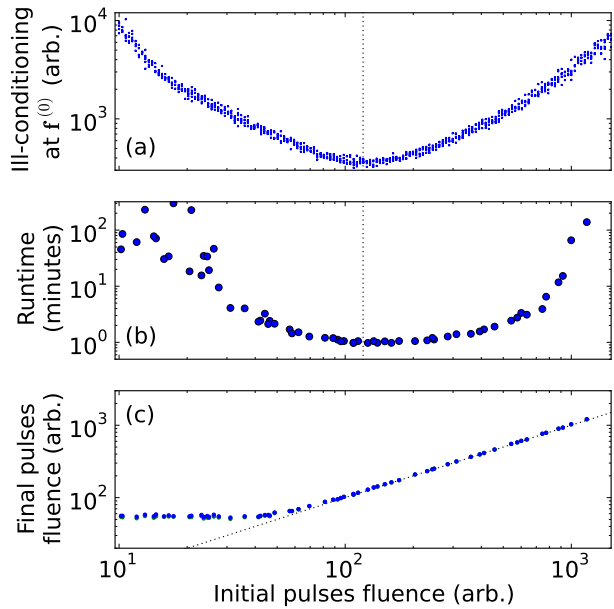}
  \caption{Direct
  analogue for our second test problem of Fig. 2 from the main text. For
  Newton-Raphson runs with initial pulses $\text{{\tmstrong{f}}}^{ \left( 0
  \right)}$ of different norms, (b) the wall time needed to reach an error
  $\varepsilon$ of $10^{- 4}$ and (c) the norm of the corresponding solution
  pulses, with a dashed `initial equals final' line. In addition, (a) the
  ill-conditioning of the Jacobian at several randomly sampled pulses of each
  norm.}
\end{figure}

As a second test problem, we can consider implementing a logical T-gate
encoded with the five physical qubit stabilizer code as described in
{\cite{nigmatullin_implementation_2009_sup}}. The underlying system is a
Heisenberg spin chain of length five, with a fixed external coupling field at
a Rabi frequency of $10$, so that $H_0$ reads
\[\sum_{n = 1}^4 \sigma_x^{\left( n \right)} \sigma_x^{\left( n + 1
\right)} + \sigma_y^{\left( n \right)} \sigma_y^{\left( n + 1 \right)} +
\sigma_z^{\left( n \right)} \sigma_z^{\left( n + 1 \right)} + 10 \sum_{n =
1}^5 \sigma_x^{\left( n \right)}\]
with an evolution time $T = 90$. This has a single control corresponding to
$H_1 = \sigma_z^{\left( 1 \right)}$
enabling the first spin to be detuned, through a local voltage which is
piecewise constant over $K = 1500$ intervals.

\end{document}